\documentclass[final,1p,times]{elsarticle} 
\usepackage{graphicx} 
\usepackage{amssymb} 
\usepackage{amsthm} 
\usepackage{lineno} 

\journal{Nuclear Physics A} 

\newcommand{\alice}{ALICE}

\newcommand{\dedx}{\ensuremath{\mathrm{d}E\mathrm{/dx}}}
\newcommand{\rphi}{\mbox{${\mathrm{r}\phi }$}}

\newcommand{\pt}{\ensuremath{p_{\mathrm{t}}}}

\newcommand {\gevc} {\mbox{\rm GeV$\kern-0.15em /\kern-0.12em c$}}
\newcommand {\tevc} {\mbox{\rm TeV$\kern-0.15em /\kern-0.12em c$}}

\newcommand {\gev} {\mbox{${\rm GeV}$}}

\newcommand {\degree} {\mbox{${^\circ}$}}

\newcommand {\mum} {\mbox{${\mu \rm m}$}}

\begin{document} 

\begin{frontmatter} 


\title{Commissioning and Prospects for Early Physics with ALICE}

\author{P.G.~Kuijer, for the \alice\ collaboration}
\address{NIKHEF, P.O.box 41882, 1009 DB  Amsterdam, Netherlands and CERN}

\begin{abstract} 
The \alice\ detector has been commissioned and is ready for taking data at the Large Hadron Collider. The first proton-proton collisions are expected in 2009. This contribution describes the current status of the detector, the results of the commissioning phase and its capabilities to contribute to the understanding of both pp and PbPb collisions
\end{abstract} 

\end{frontmatter} 



\section{INTRODUCTION}\label{sec:introduction}
\alice\ is a general-purpose heavy-ion experiment designed to study the physics of strongly interacting matter and the quark-gluon plasma in nucleus-nucleus collisions at the LHC. The \alice\ detector~\cite{ref:PPRI} is designed to deal with  large particle multiplicities, dN/dy up to 8000, well above the multiplicities expected for PbPb collisions at LHC energies. 

The \alice\ collaboration will also study collisions of lower-mass ions and protons.  The pp collisions will primarily provide reference data for the nucleus-nucleus collisions but in addition a number of genuine pp physics studies will be done.

\begin{figure}[ht]
\centering
\includegraphics[width=0.35\textwidth]{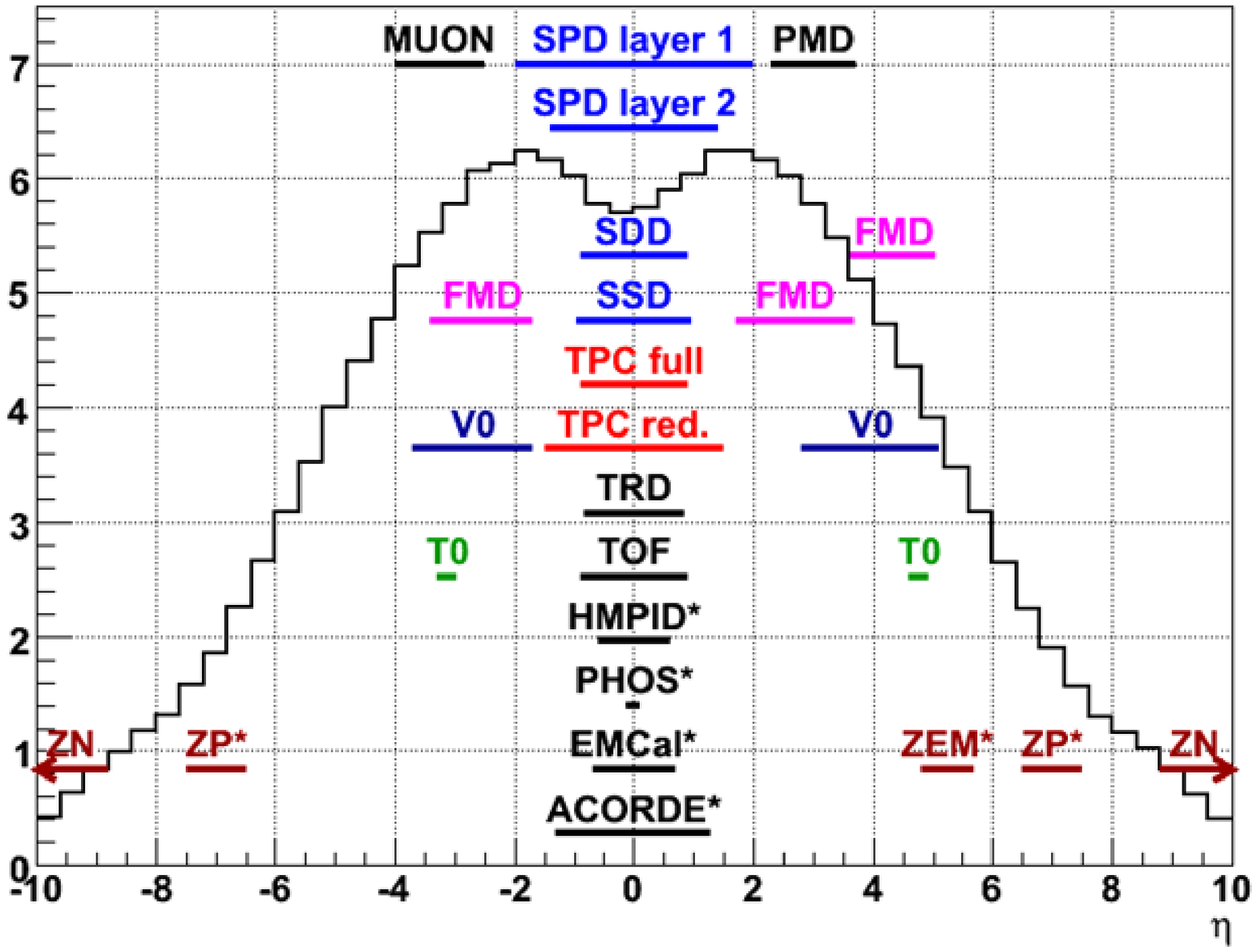}
\includegraphics[width=0.60\textwidth]{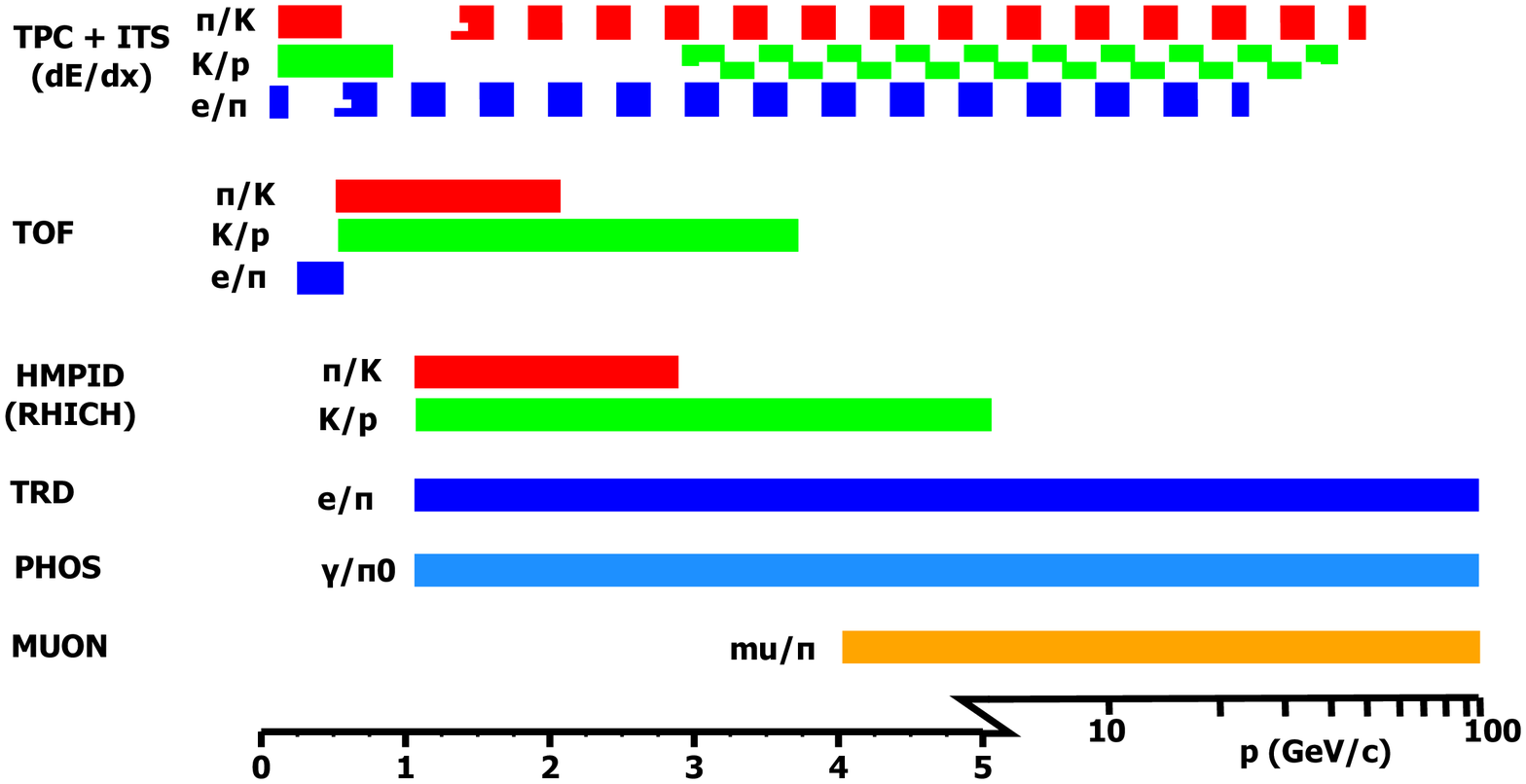}
\caption{\alice\ Pseudo rapidity coverage of the \alice\ detectors (left panel),  and the particle identification capabilities of the detectors right panel).} 
\label{fig:properties}
\end{figure}
The main components of the \alice\ detector are a central tracking and particle identification system covering the pseudo rapidity range $-0.9\leq\eta\leq0.9$, a muon spectrometer covering $-4.0\leq\eta\leq-2.4$, a forward multiplicity 
detector and a zero-degree calorimeter. An overview of the $\eta$ coverage of the \alice\ systems is shown in the left panel of Figure~\ref{fig:properties}. 

The transverse momentum cut-off is only 0.1~\gevc\ due to the extreme minimisation of the material budget of the inner tracking system. On the other hand the \pt\ measurement reaches up to 50~\gevc\ with full particle identification capabilities at mid-rapidity. 

The central part is embedded in the large L3 solenoidal magnet which provides a field of 0.5~T. The tracking system is subdivided in the inner tracking system (ITS) using different kinds of silicon detectors and a large time projection chamber (TPC). 
The inner tracking system consists of six concentric cylindrical layers of detectors. From inside out the inner tracking system consists of two layers of silicon pixel detectors, two layers of silicon drift detectors and two layers of silicon strip detectors. The TPC consists of two 2.5~m long drift volumes separated by a central cathode. The inner tracking system determines the vertex resolution of the system while the TPC essentially defines the momentum resolution. 
 
The transition radiation detector and the time of flight (TOF) array cover the full azimuthal angle. Additional detectors with partial coverage of the central barrel are the photon spectrometer, the high momentum particle identification detector and the electromagnetic calorimeter.
An overview of the \alice\ particle identification capabilities is shown in the right panel of Figure~\ref{fig:properties}. 
Details of the detector design and the expected performance are described in~\cite{ref:PPRI}.

Read-out of the muon spectrometer is triggered by the muon trigger chambers. The central tracking system is triggered by a combination of dedicated trigger detectors (T0, V0) and a trigger derived from the silicon pixel detectors (SPD) in the innermost layer of the tracking system. Each chip in the SPD front-end system outputs a ``fast-or'' signal indicating that at least one pixel has fired. These signals are combined in a global multiplicity trigger which can be adjusted to select central events in PbPb collisions. However, since the noise in pixel detectors is intrinsically low, the same system was used to trigger on single cosmic muons for commissioning purposes.

The hardware triggering system is complemented by the high level trigger (HLT) system which uses a large processor farm to select or tag events after read-out of the hardware. About 30\% of this system was installed during the detector commissionng phase. In addition to its event selection capabilities this system provides a data quality monitoring facility.

The experiment is ready for data taking with both proton and heavy-ion beams.  Partially installed are TRD (25\%, completed 2010), PHOS (60\%, completed 2010) and EMCAL (completed 2011). The central tracking system, the time of flight system and the forward muon spectrometer are fully installed. Also the data acquisition system (DAQ) and the detector control system (DCS) are fully installed. Therefore at start-up \alice\ has full hadron and muon identification capabilities and partial photon and electron identification capabilities.
\section{COMMISSIONING}\label{sec:commissioning}
All installed detectors have been tested concurrently for noise performance in the L3 magnetic field. Cabling, grounding and power supply issues have been resolved. During 2007/2008 three cosmic data taking periods with runs involving all detectors 24~hours a day were taken.  The first run of 12 days in December 2007 was mainly used to debug the global and detector specific data taking systems. The second run of five weeks (February-March 2008) was used to verify that all systems could efficiently operate without mutual interference. The majority of the commissioning data was collected in the third run which started in May 2008 and lasted until the foreseen start-up of the LHC in October.
\begin{figure}[ht]
\centering
\includegraphics[width=0.47\textwidth]{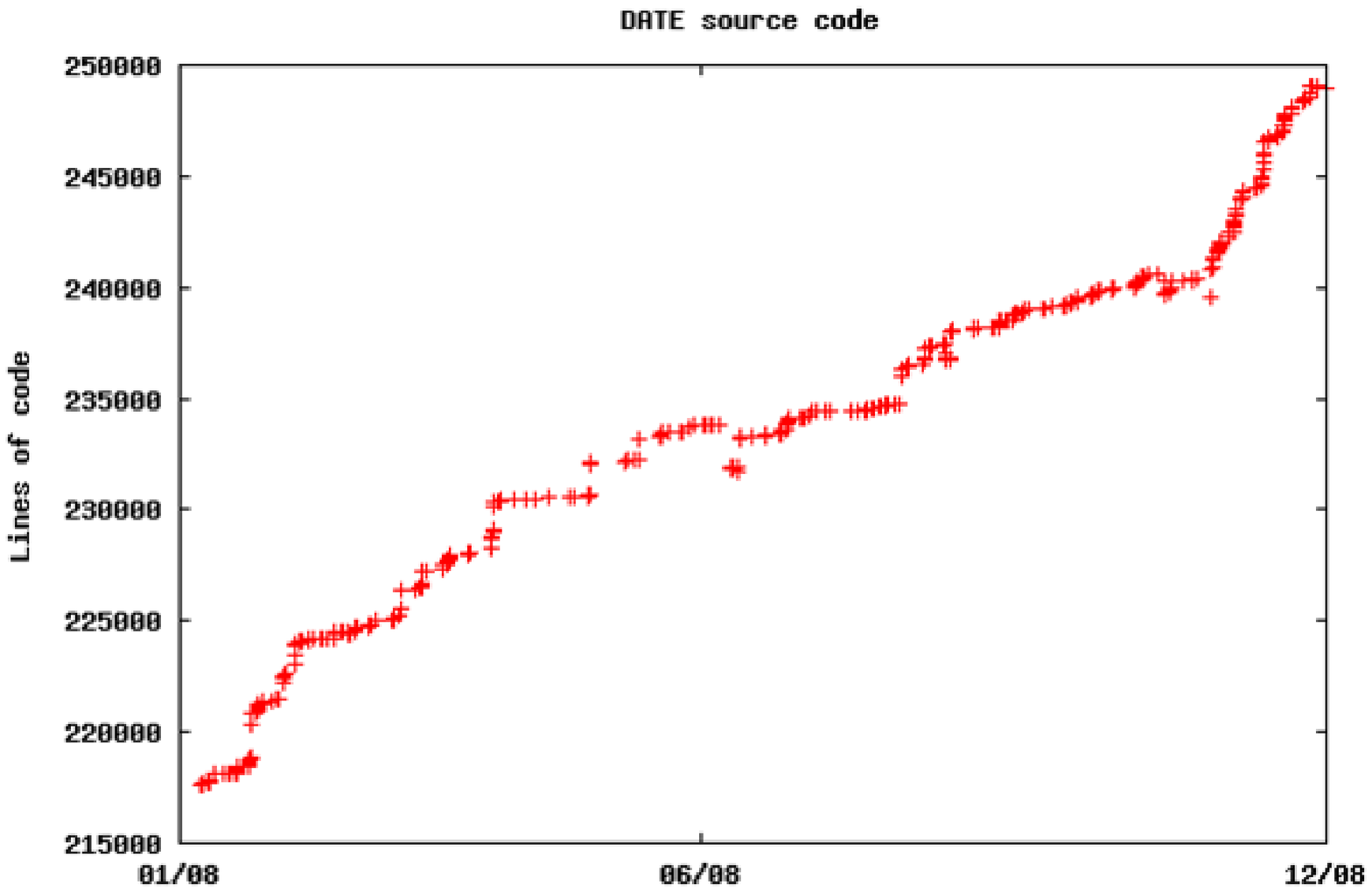}
\includegraphics[width=0.48\textwidth]{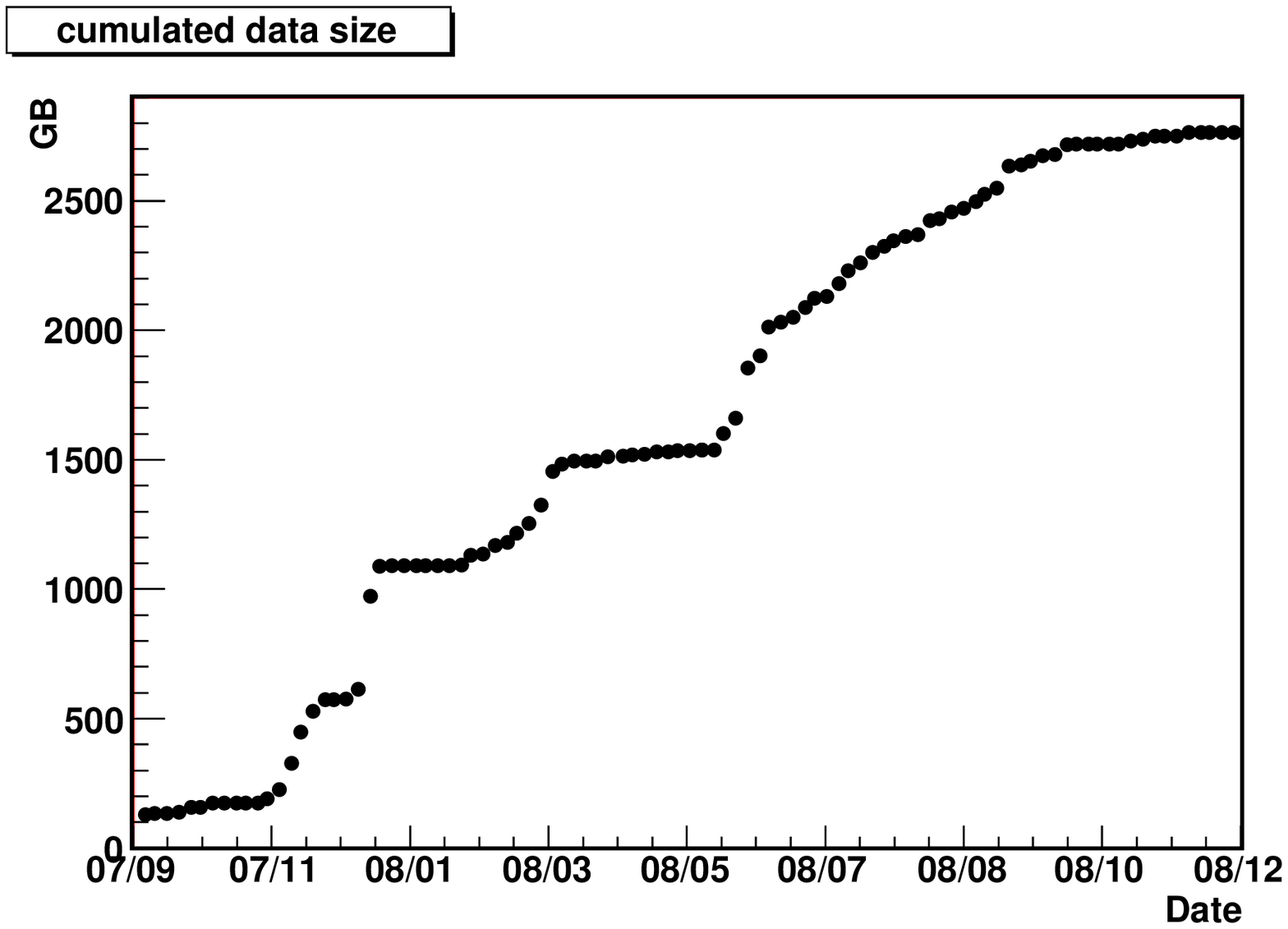}
\caption[]{Left panel: Online source code size as a function of time in 2008. The code has grown by about 15\% with a large increase in the activity near the end of the datataking period. Right panel: Amount of data accumulated by the DAQ system.} 
\label{fig:online}
\end{figure}
Using the silicon pixel trigger, adjusted to trigger on single particles, a sample of muons crossing the entire tracking system has been collected in 2008. 
In addition the dedicated cosmic trigger system (ACORDE) was used to collect a large sample of single and multiple cosmic muon events. The muon spectrometer was triggered using its own muon trigger system. The total read-out rate was about 100~Hz during several months, thus testing both the data acquisition system and the offline reconstruction software thoroughly. The availability of real data from the detectors has stimulated the implementation of additional features in the online software, as illustrated in Figure~\ref{fig:online}. The code has grown by about 15\% with a large increase in the activity near the end of the datataking period. In addition valuable experience in operating all systems was gained allowing for efficient datataking when the LHC starts delivering beams. In 515~days of datataking a total of about 3~PB of data was read-out of which 350~TB was recorded to tape, see Figure~\ref{fig:online}.

The muon spectrometer has collected a small sample of nearly horizontally moving cosmic muons which allowed to check the trigger system and gave a first indication of the internal alignment of the chambers.
The cosmic data provided an opportunity to align the central tracking systems and partially calibrate them. For the smaller detectors beams will be needed to calibrate and align because they are in an unfavourable orientation with respect to cosmic muons or because their area is too small to collect a useful sample in an acceptable time. The following paragraphs describe the results obtained for the central tracking systems. For further details on the ITS commissioning results see also~\cite{ref:prino}.

\begin{figure}[ht]
\centering
\includegraphics[width=0.35\textwidth]{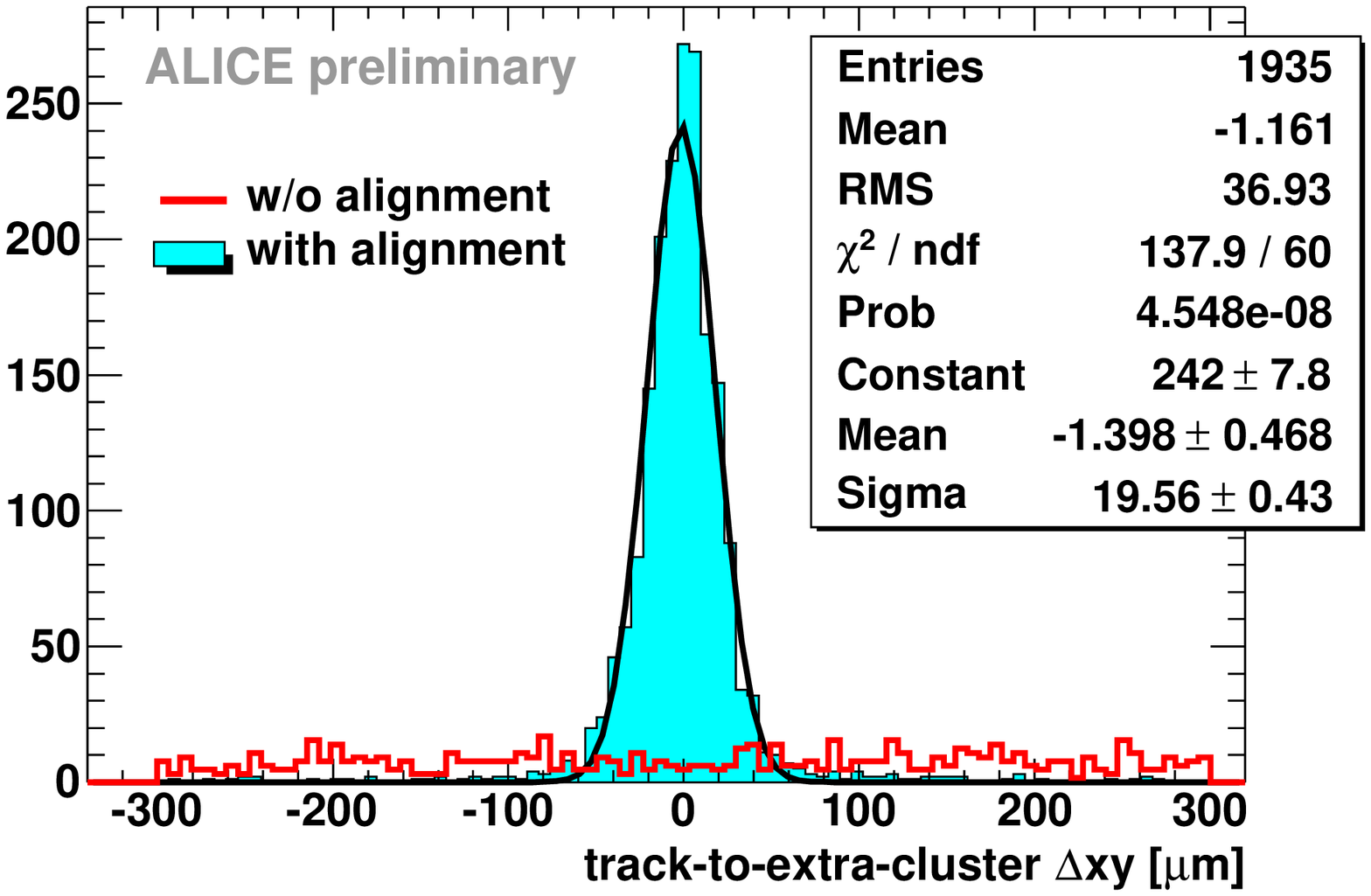}
\includegraphics[width=0.25\textwidth]{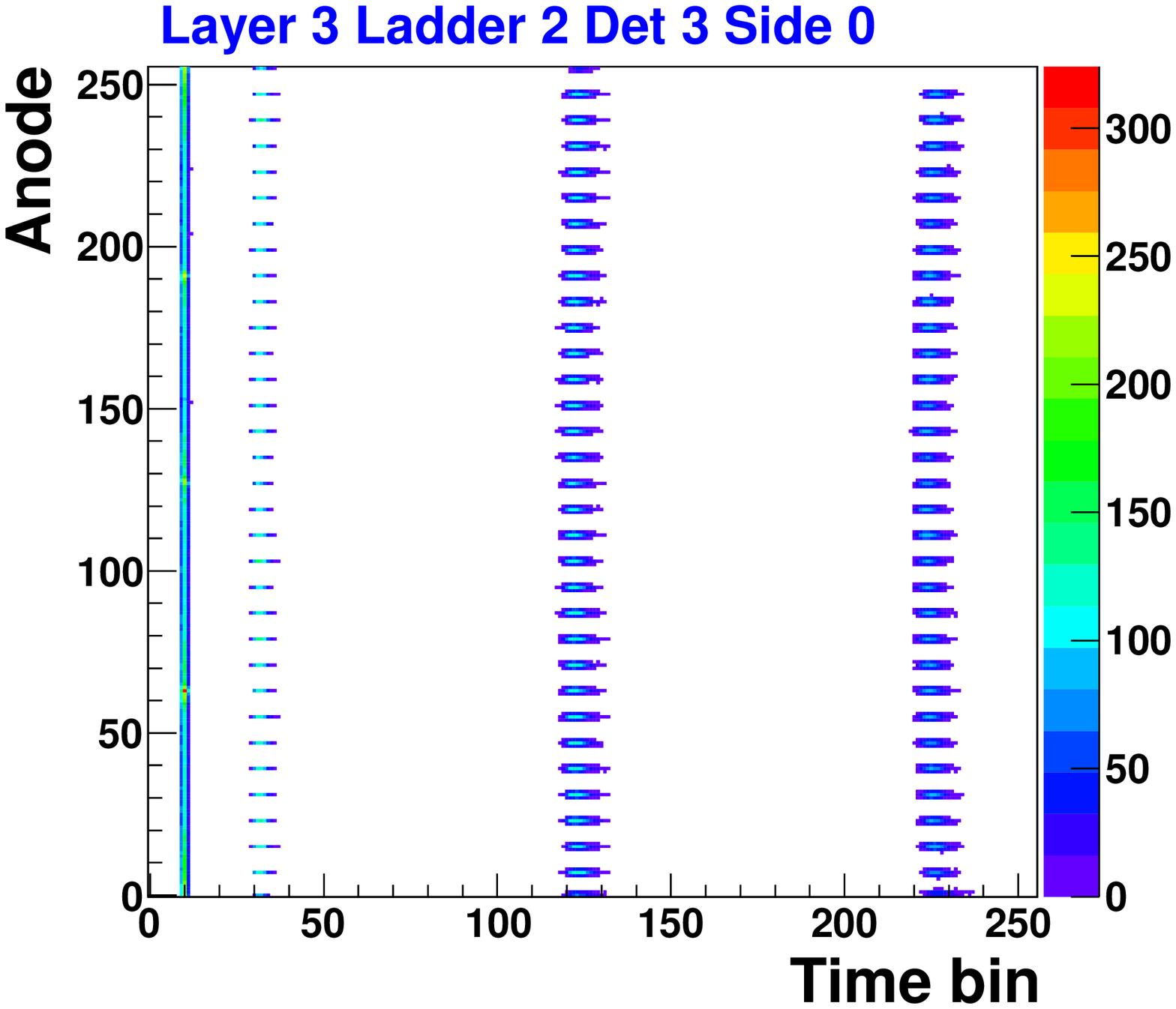}
\includegraphics[width=0.35\textwidth]{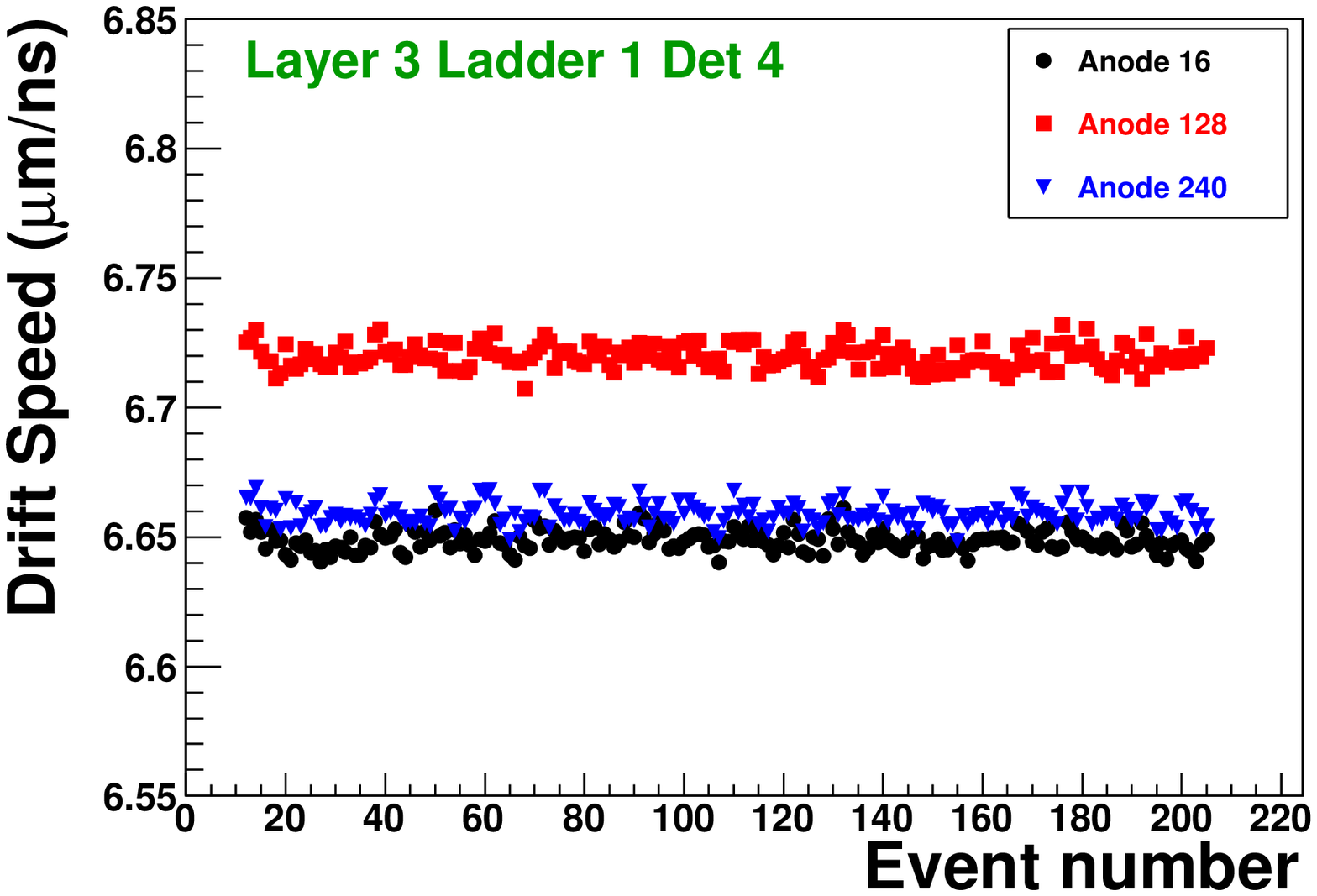}
\includegraphics[width=0.30\textwidth]{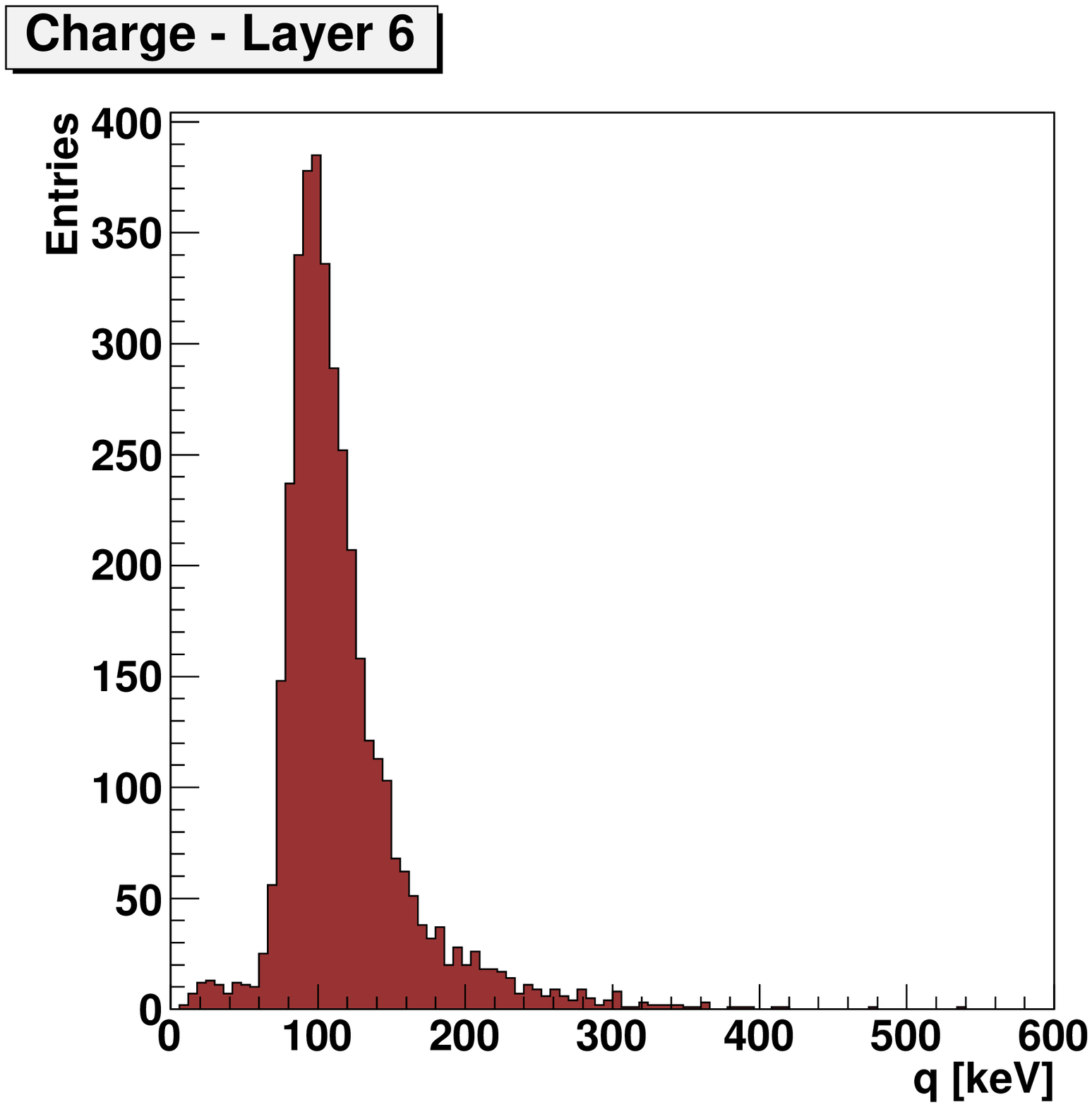}
\includegraphics[width=0.30\textwidth]{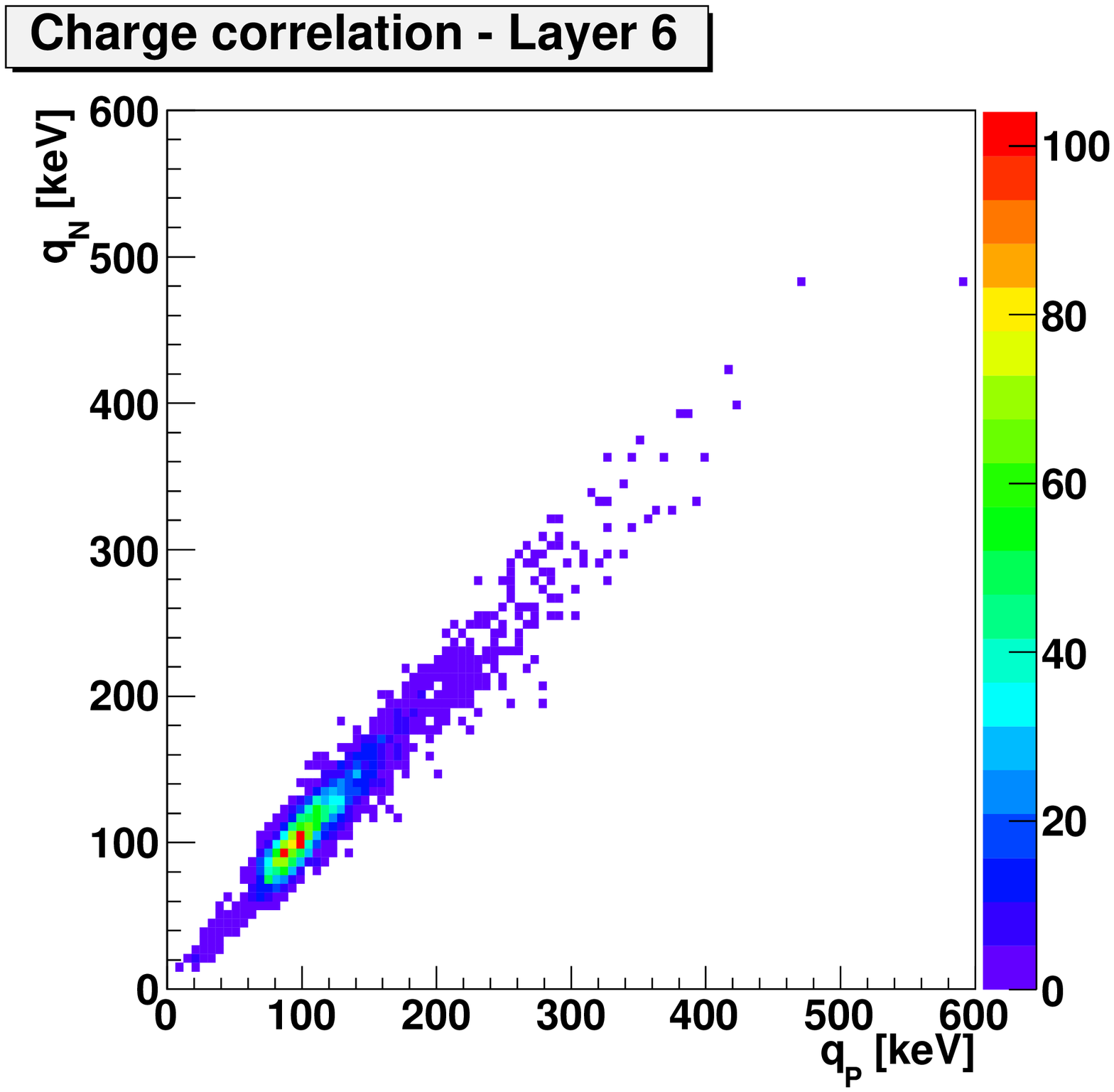}
\caption[]{Calibration and alignment results for the inner tracking system. Upper left: track to track distance before and after alignment. Upper middle panel: drift speed map of a silicon drift detector. Upper right panel: Drift speed as a function of time for a silicon drift detector. Lower left panel: measured energy loss in a silicon strip detector after calibration. Lower right panel: Measured charge correlation between the two sides of a silicon strip detector} 
\label{fig:its}
\end{figure}
In case of a single muon crossing the inner layers of the tracking system the reconstruction software finds two tracks because it searches for tracks coming from the interaction region. The distance between the two tracks provides a measure of the vertex resolution. The result before and after alignment of the individual layers of the inner tracking system is show in Figure~\ref{fig:its} (upper left panel). The vertex resolution depends mostly on the alignment of the inner (pixel) layers. The estimated vertex resolution is 40~\mum, close to the design value.

The silicon drift detectors measure the \rphi\ coordinate by measuring the arrival time of the electrons on the anodes. However, the drift speed depends strongly on the temperature. Therefore the \alice\ silicon drift detectors have integrated electron injection devices which allow to monitor the drift speed continuously. During the cosmic run the drift speed was measured as a function of the position on the detectors and of time. Figure~\ref{fig:its} shows the result of the drift speed distribution on a detector (upper middle panel) and the speed as a function of time (upper right panel). The drift speed is stable during many hours of operation as a consequence of the stable thermal conditions in the inner tracking system. This allows an effective reconstruction of the position using the measured drift speed map. In addition the drift detectors were aligned and the 
\dedx\ measurement was calibrated.

The outer two layers of the ITS, consisting of silicon strip detectors were aligned using the cosmic muon tracks. In addition to the position measurements these detectors provide a \dedx\ measurement. An example of the response is shown in Figure~\ref{fig:its} (lower left panel) demonstrating that a signal to noise ratio of about~40 is achieved. Because the silicon strip detectors in \alice\ are double sided they provide both the \rphi\ and z coordinates in a single detector. However, although the stereo angle between the strips on each side is relatively small, ambiguities may arise if two particle cross the detector close to each other. A significant fraction of these ambiguities can be resolved because the charge measured on both sides of the detector is the same apart from the noise in the respective channels. An example of the correlation of the charge measurements on both sides of the same detector is show in Figure~\ref{fig:its} (lower right panel).

\begin{figure}[ht]
\centering
\includegraphics[width=0.26\textwidth]{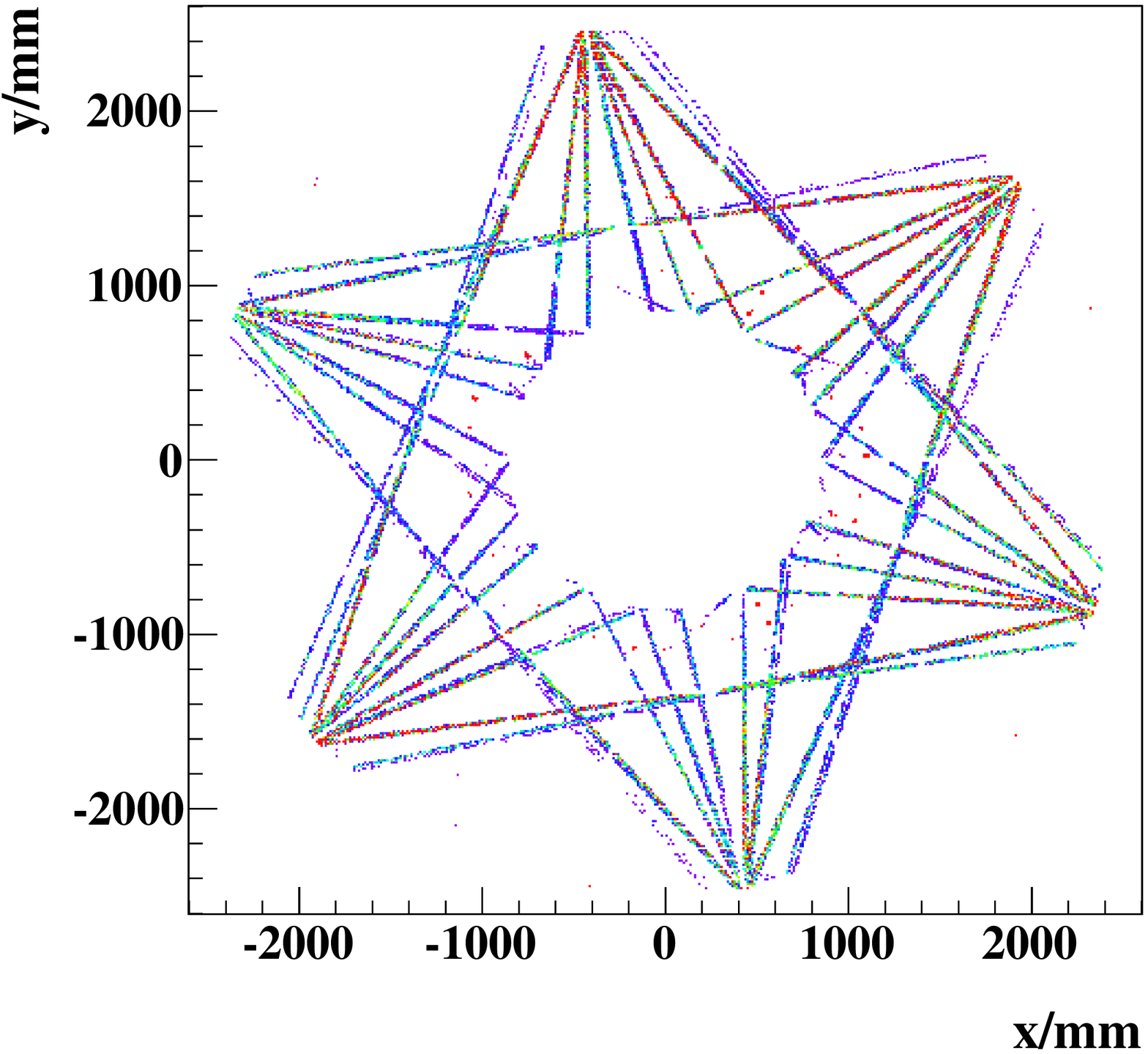}
\includegraphics[width=0.26\textwidth]{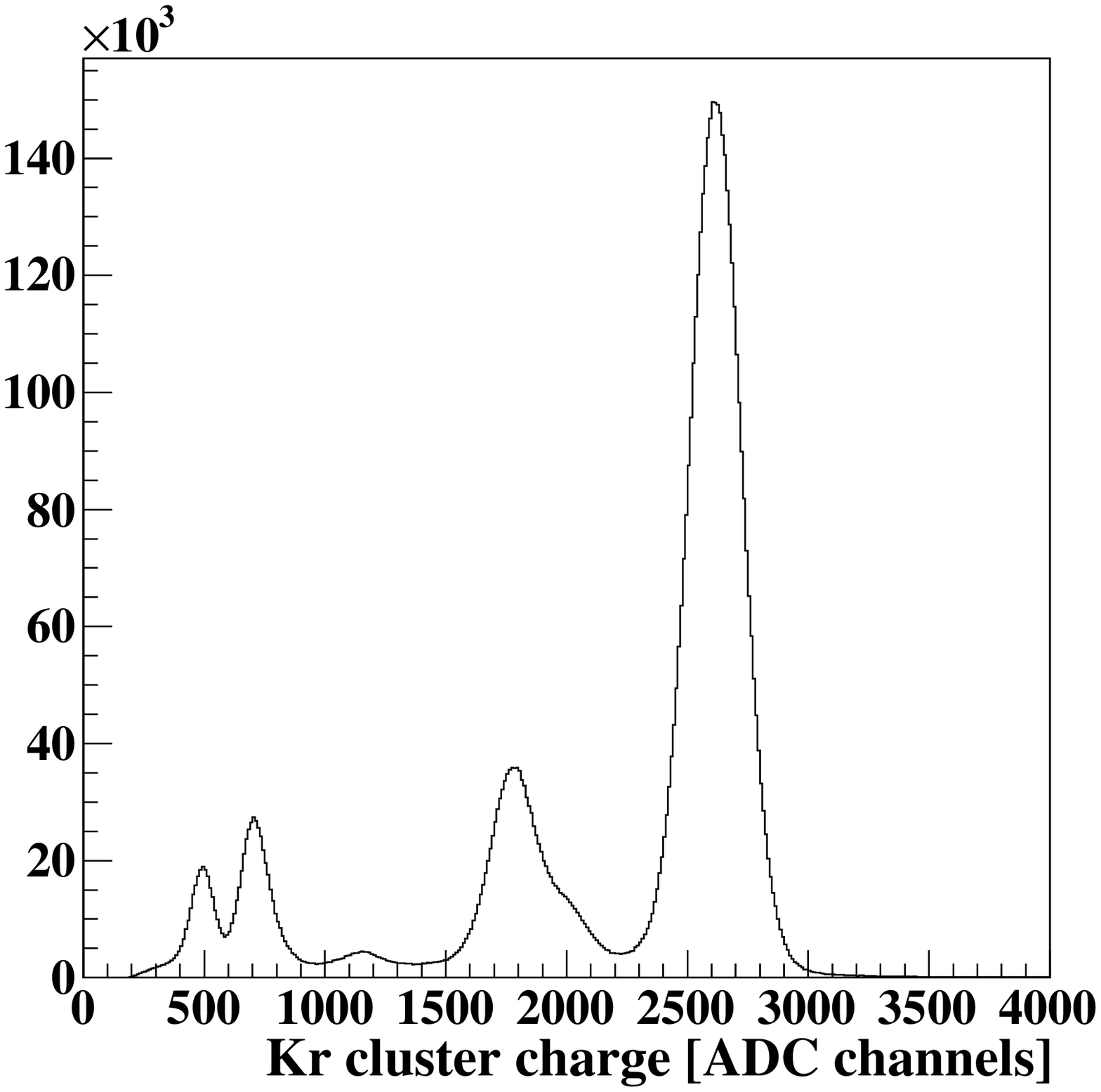}
\includegraphics[width=0.32\textwidth]{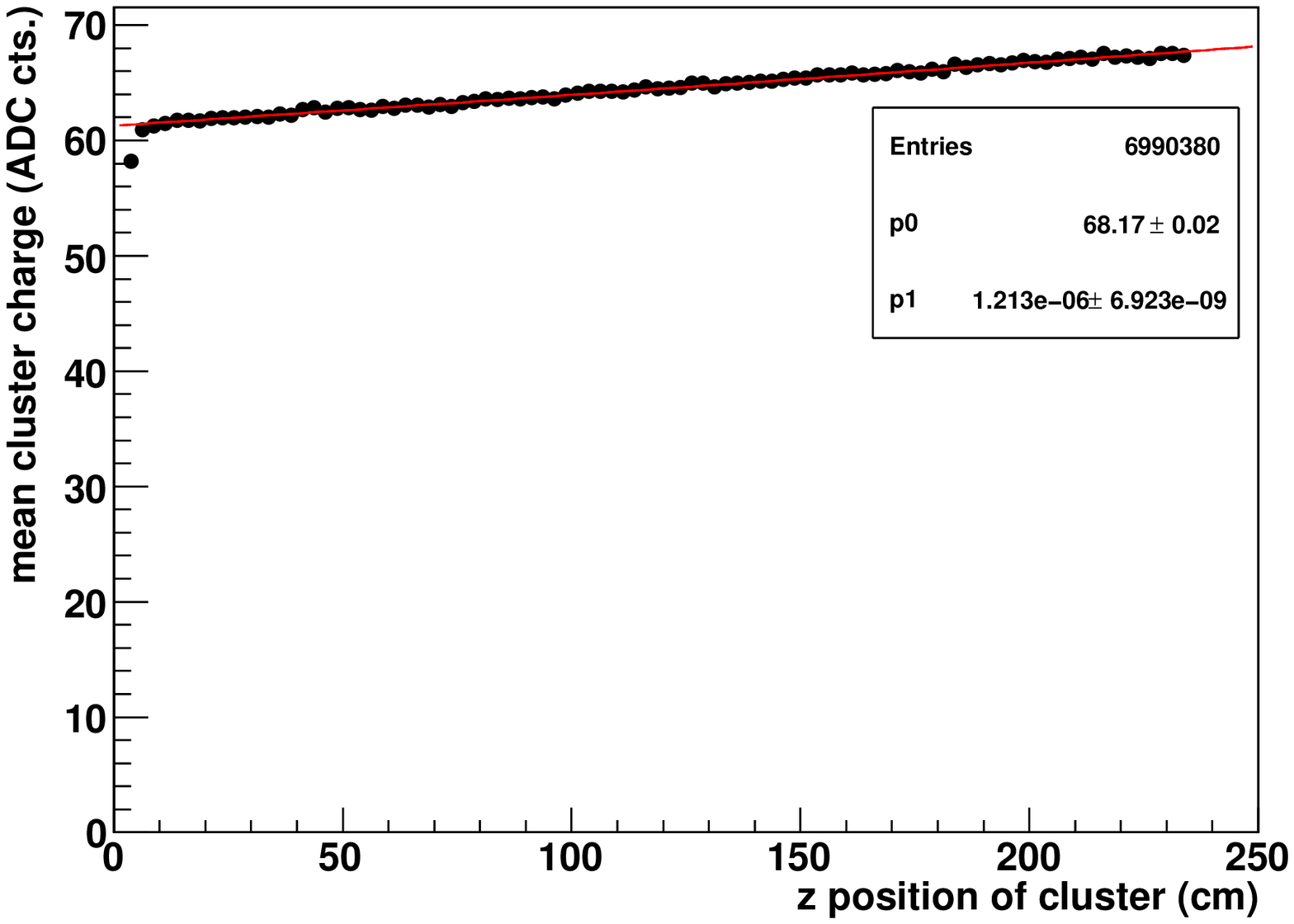}
\includegraphics[width=0.35\textwidth]{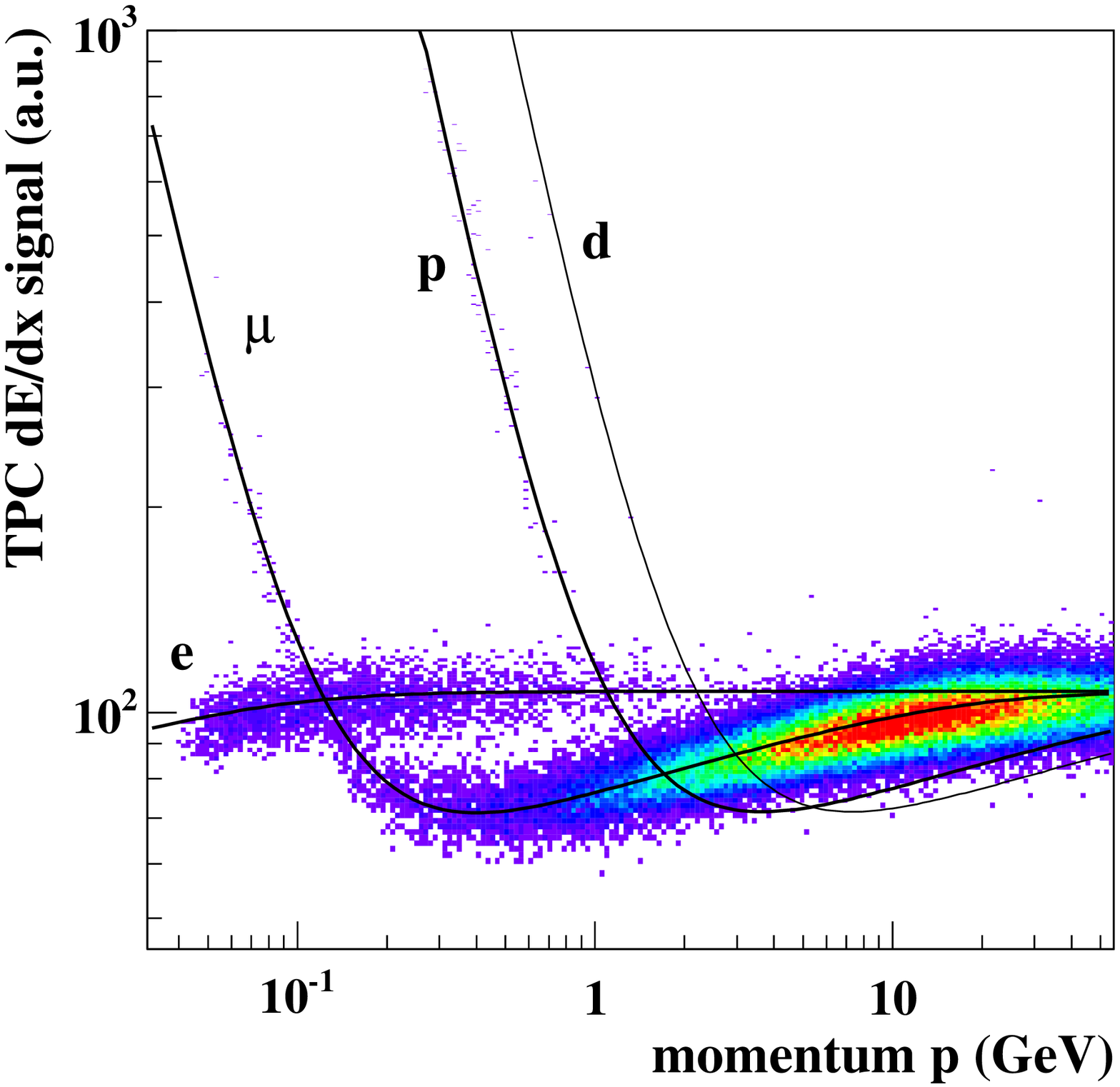}
\includegraphics[width=0.30\textwidth]{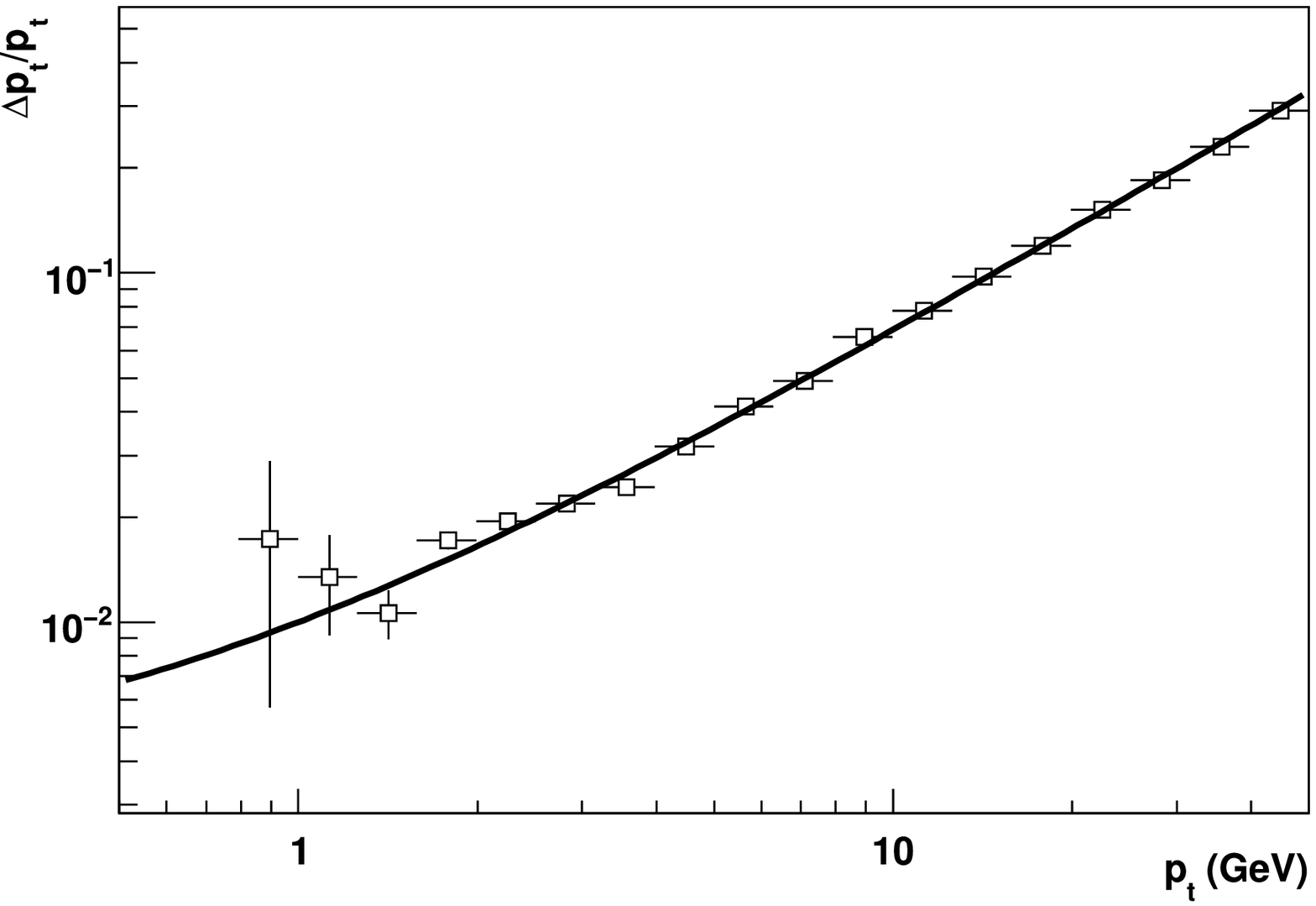}
\caption[]{\alice\ particle identification in the TPC from cosmic shower events} 
\label{fig:tpc}
\end{figure}
The time projection chamber has an integrated laser calibration system in order to measure the drift speed. Figure~\ref{fig:tpc}, upper left panel, shows the measurement of the ionisation trails of the laser beams. In order to calibrate the \dedx\ measurement in the TPC a radioactive gas was injected in the drift volume. The measured energy spectrum of the Kr decays, see the upper middle panel of Figure~\ref{fig:tpc}, shows the excellent energy resolution of the TPC. Due to electron attachment and diffusion the measured charge depends on the position of the ionisation in the drift volume. The upper right panel of Figure~\ref{fig:tpc} shows the amplitude as a function of the position. Both the \dedx\ measurement and the drift speed must be corrected for the gas pressure and temperature which are also continuously monitored. The extensive commissioning period provided ample data to verify all necessary corrections. 

Due to its large volume the TPC could collect more than 10M cosmic muon and particle shower events. This large sample allowed to test the particle identification capabilities of the TPC as shown in Figure~\ref{fig:tpc} lower left panel. A clear separation of electrons, muons and protons is demonstrated at momenta up to 3~\gevc. The \dedx\ resolution, without a thorough calibration, is already 5.7\%, very close to the design value of 5.5\%.

The momentum resolution was studied by separating the cosmic tracks into two halves and comparing the reconstructed momenta. Already after the first attempt to align and calibrate a momentum resolution of 6\% at 10~\gevc\ is achieved, reassuringly close to the design value of 4.5\%~\cite{ref:wiechula}. 

\begin{figure}[ht]
\centering
\includegraphics[width=0.55\textwidth]{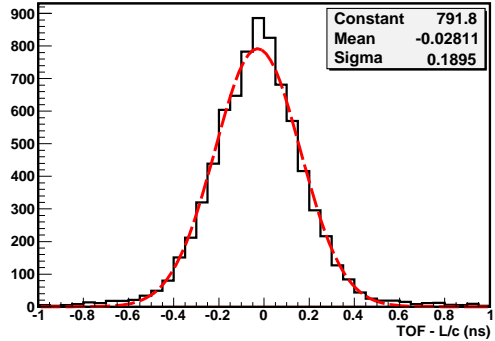}
\caption[]{Time of flight distribution measured using cosmic muons. The time resolution of a single module is found by dividing the width of the distribution by $\sqrt{2}$, i.e. 130~ps.} 
\label{fig:tof}
\end{figure}
For the time of flight (TOF) system a special purpose cosmic trigger, using the signals from the TOF system itself, was developed to collect a sample of cosmic muon events crossing two of the TOF modules. The time projection chamber was used to measure the momentum of the muon and the track length between the two TOF modules. Since all muons measured this way are relativistic the time difference between the two modules measured by the TOF can be compared to the time difference calculated from the track length. The time resolution of a single module is then found by dividing the width of the distribution by $\sqrt{2}$. Thus the time resolution of the TOF system, even with the current preliminary calibration, was found to be 130~ps, see also Figure~\ref{fig:tof}, showing that also the TOF performs close to its design specifications.

\begin{figure}[ht]
\centering
\includegraphics[width=0.28\textwidth]{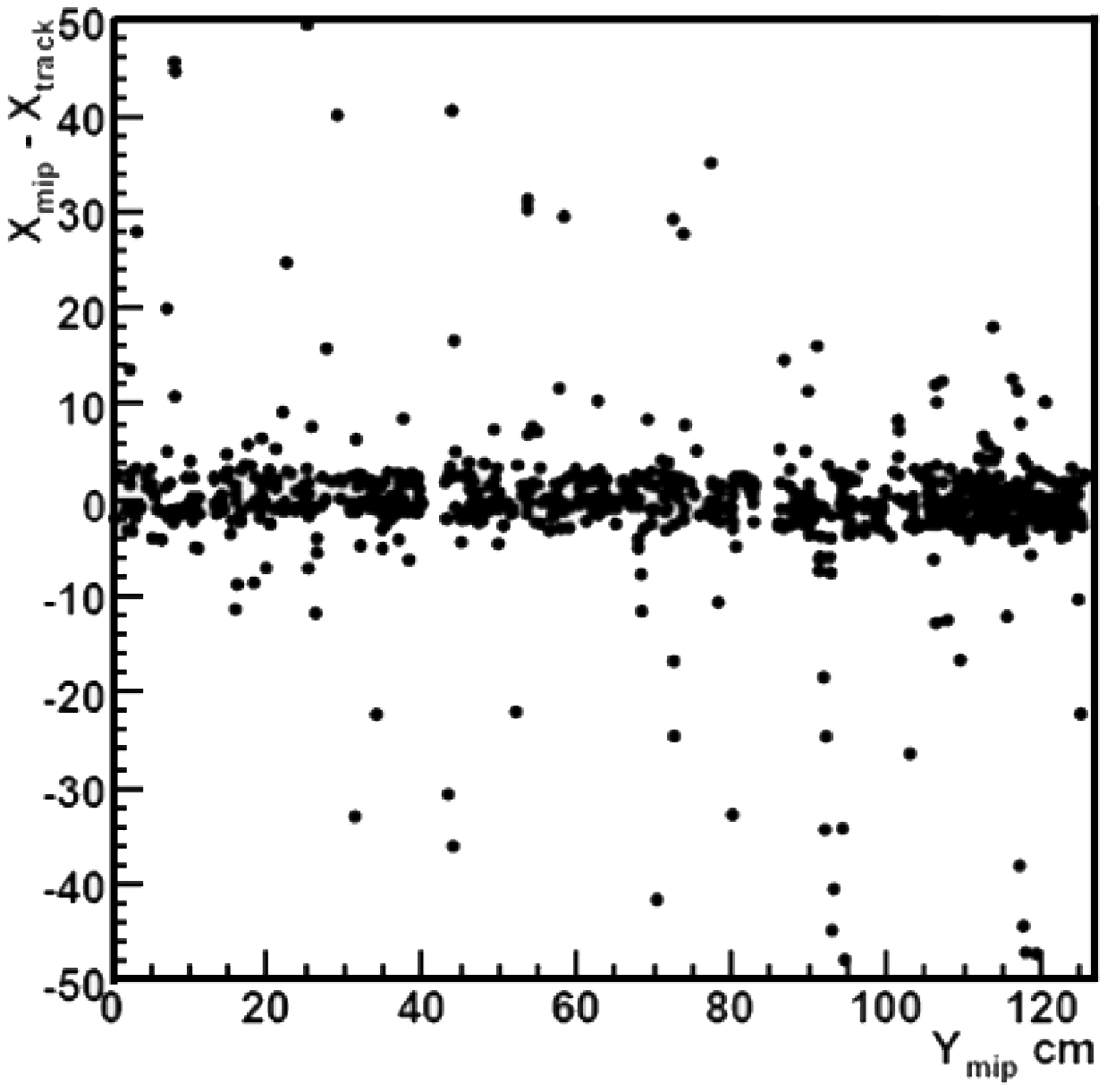}
\includegraphics[width=0.37\textwidth]{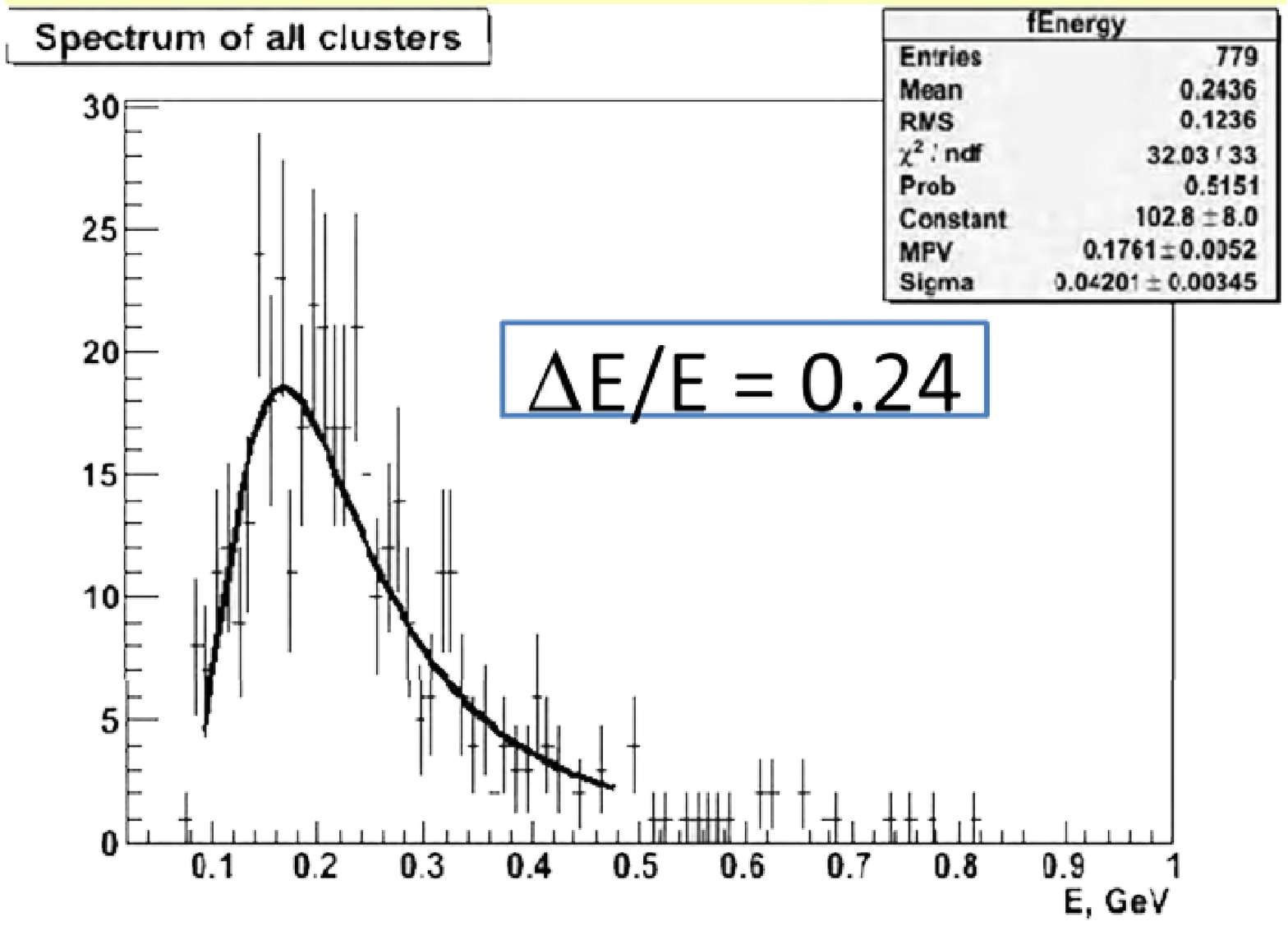}
\includegraphics[width=0.30\textwidth]{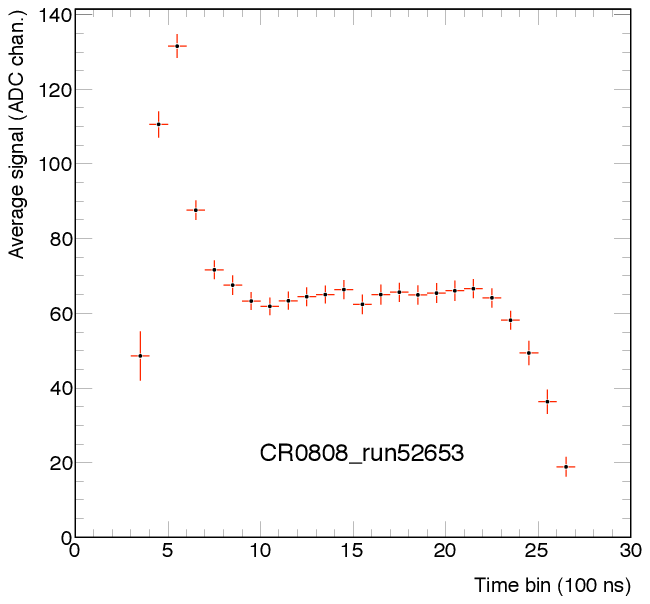}
\caption[]{Left panel: HMP-TPC track matching. Centre panel: PHO energy resolution for cosmic muons. Right panel: First TRD signals.} 
\label{fig:other}
\end{figure}
The high momentum particle identification system (HMPID), the transition radiation detector (TRD) and the photon spectrometer (PHOS) are placed in unfavourable positions for detecting cosmic muons. Nevertheless small samples of data were collected with which the proper functioning of the detectors could be verified. Figure~\ref{fig:other}, left panel, shows the measurement of the track matching precision between the TPC and the HMPID without a precise alignment~\cite{ref:volpe}. The right panel of the same figure shows the first observed TRD signals~\cite{ref:kweon}. The PHOS system is not yet completed but one module was operated at room temperature, instead of -20\degree\ as foreseen in the design. Nevertheless even at room temperature an energy resolution of dE/E of 0.24 was obtained, see Figure~\ref{fig:other} centre panel.

\begin{figure}[ht]
\centering
\includegraphics[width=0.35\textwidth]{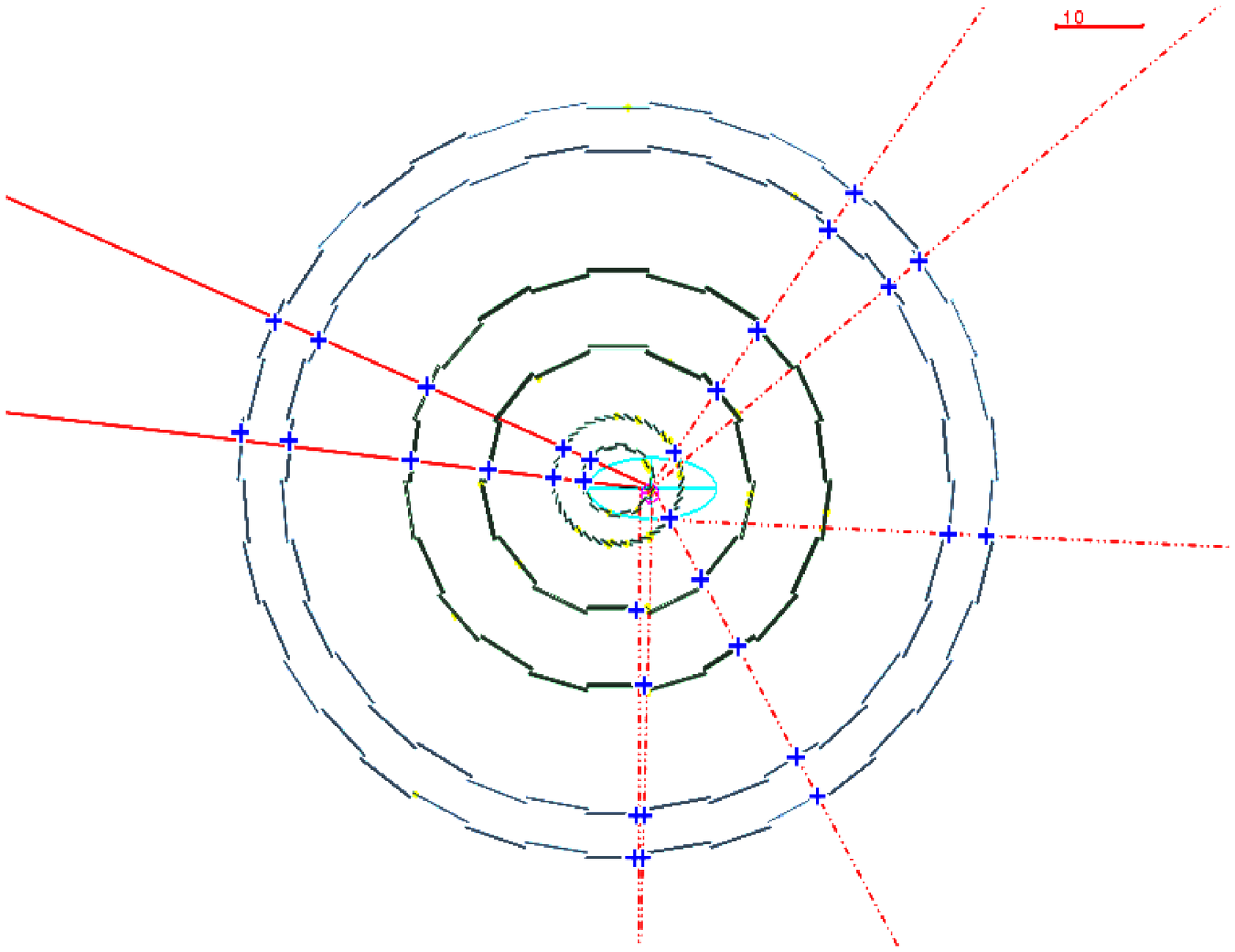}
\caption[]{First event event seen during an LHC injection test} 
\label{fig:first}
\end{figure}
In August and September 2008 the first protons were injected into the LHC. While the particles extracted from the SPS were dumped at the end of the extraction line or at a collimator in the LHC the relative timing of the \alice\ trigger detectors could be studied. The time differences are sufficiently small to allow equalisation in the central trigger processor so that the individual triggers can be combined into coincident trigger conditions. The particle showers emitted from the dump at the end of the extraction line provided excellent test signals to compare the particle densities measured by the forward multiplicity detector (FMD), the silicon pixel detectors, the V0 trigger counters and the muon spectrometer. 

When the first particles were sent through the \alice\ experiment into the next section of the LHC the first beam interaction was observed in the central tracking system. Figure~\ref{fig:first} shows the reconstructed event containing an interaction between a stray particle from the beam and a silicon pixel detector in the innermost layer of the inner tracking system. 

\section{Prospects for early physics}\label{sec:pp}
The LHC accelerator is expected to resume operations in 2009. At start-up some pp~collisions at 900~\gev will be provided, followed by a pp run at the highest possible LHC energy. 
PbPb collisions are expected in 2010. 

The \alice\ detector was optimized for the high multiplicity environment resulting from heavy ion collisions, see Section~\ref{sec:introduction}. This also gives it unique capabilities for the measurement of proton-proton collisions. 

The first proton-proton collisions at 900~\gev, expected in 2009, will allow the experiment to compare reults with previous experiments. 
During the following high energy pp run \alice\ will collect a large sample of minimum bias events providing the multiplicity and \pt-distributions as well as important reference data for the heavy-ion programme. In addition, the unique capabilities of \alice\ will allow it to contribute significantly to the measurement of the baryon transport mechanism. The measurement of the charm cross section will provide important input to the pp QCD physics. 

The first 10$^{5}$ PbPb events, collected in less than 1~day, will provide global event properties such as the multiplicity distribution, the rapidity density and elliptic flow. The source characteristics (particle spectra, resonances, differential flow and interferometry) can be extracted from the first 10$^{6}$ events (1 week). In the first year of PbPb collisions \alice\ will collect 10$^{7}$ minimum bias events and an equal number of central collisions (5\% most central), allowing the study of charmonium production, jet quenching and heavy-flavour energy loss in the medium. These measurements will provide information on the bulk properties of the medium, such as the energy density, temperature, pressure, heat capacity/entropy, viscosity, speed of sound and opacity.

For a detailed description of the \alice\ physics capabilities
see~\cite{ref:PPRII,ref:malek,ref:kharlov,ref:dainese}. 

\section{Summary}\label{sec:summary}
All installed detector systems were debugged and commissioned during 2008. Partially installed are the TRD (25\%, to be completed in 2010), the PHOS (60\%, to be cpompleted in 2010) and the EMCAL (to be completed in 2011). All other systems are fully installed. The central tracking systems have been tested and calibrated using cosmic muons. The performance is shown to be close to the design values for the silicon detectors, the time projection chamber and the time of flight system. The data acquisition system and the offline software have been tested with more than 350~TB of data from cosmic ray events.

The \alice\ experiment is ready to take data with pp and PbPb collisions to be provided by the LHC accelerator. 




\begin{thebibliography}{00}
\bibitem{ref:prino}
F. Prino for the \alice\ collaboration, ``Results from commissioning of the 
\alice\ ITS with cosmics'', these proceedings.
\end{thebibliography}
\end{document}